\documentclass[11pt]{article}
\usepackage[left=1in,top=1in,right=1in,bottom=1in]{geometry}
\usepackage{times}

\usepackage{amsmath}
\usepackage{amssymb}
\usepackage{amsthm}
\usepackage{graphicx}
\usepackage{verbatim}
\usepackage{url}
\usepackage[plain]{algorithm}
\usepackage{algorithmic}

\allowdisplaybreaks

\usepackage{siunitx}
\usepackage[T1]{fontenc}
\usepackage{booktabs}
\usepackage{array}
\usepackage{microtype}
\usepackage{etoolbox}

\newtheorem{theorem}{Theorem}
\newtheorem{lemma}{Lemma}

\newtheorem{definition}{Definition}

\newcommand{\eps}{\varepsilon}
\newcommand{\calH}{\mathcal{H}}
\newcommand{\calQ}{\mathcal{Q}}
\newcommand{\conv}{\operatorname{conv}}
\newcommand{\rank}{\operatorname{rank}}
\newcommand{\ord}{\operatorname{ord}_0}
\newcommand{\lc}{\operatorname{lc}_0}

\newcommand{\R}{\mathbb{R}}
\newcommand{\Z}{\mathbb{Z}}

\allowdisplaybreaks

\begin{document}

\author{
Sam Ganzfried\\
Ganzfried Research, Cornell University\\
\texttt{sam.ganzfried@gmail.com}
}

\date{\vspace{-5ex}}

\title{Polynomial-Time Computation of Extensive-Form Proper Equilibria\ in Two-Player Zero-Sum Games}

\maketitle

\begin{abstract}
We give a deterministic polynomial-time algorithm for computing an exact
extensive-form proper equilibrium in a finite two-player zero-sum
extensive-form game with perfect recall and rational data. Properness is
understood in the \emph{agent normal form}: an action at an information set
is penalized relative to a better action while all other local behavioral
choices, including the player's own future actions, remain fixed.
Our algorithm assigns each information set a geometric permutahedron of
allowable action probabilities. A scaled doubly stochastic representation
embeds these restrictions in a polynomial-size sequence-form polytope;
a saddle point then satisfies the local $\eps$-properness inequalities.
To obtain an exact equilibrium rather than a solution at an arbitrarily
small numerical perturbation, we establish sign stability of a parametric
linear system, construct a persistent feasible basis, and extract the
limits of \emph{behavioral ratios}. The resulting probabilities are rational
and have polynomial encoding length, including at information sets whose
realization probabilities vanish.
\end{abstract}

\begin{table*}[t]
\centering
\small
\setlength{\tabcolsep}{4pt}
\renewcommand{\arraystretch}{1.20}
\begin{tabular}{@{}>{\raggedright\arraybackslash}p{0.270\textwidth}>{\raggedright\arraybackslash}p{0.432\textwidth}>{\raggedright\arraybackslash\leavevmode}p{0.260\textwidth}@{}}
\toprule
\textbf{Refinement} & \textbf{Main distinction} & \textbf{Polynomial-time result}\\
\midrule
Normal-form trembling-hand perfect (NFPE) & Trembles over complete contingent plans & Miltersen and S{\o}rensen (2006)~\cite{Miltersen06:Computing}\\
Sequential equilibrium (SE) & Consistent beliefs and sequential rationality & Miltersen and S{\o}rensen (2006)~\cite{Miltersen06:Computing}\\
Normal-form proper (NFPropE) & Payoff-ranked trembles over full contingent plans & Miltersen and S{\o}rensen (2008)~\cite{Miltersen08:Fast}\\
Quasi-perfect (QPE) & Best responses do not anticipate own future mistakes & Miltersen and S{\o}rensen (2010)~\cite{Miltersen10:Computing}\\
Extensive-form trembling-hand perfect (EFPE) & Trembles at individual information sets (agent form) & Farina and Gatti (2017)~\cite{Farina17:Extensive}\\
One-sided quasi-perfect (OSQPE) & Only the potentially imperfect opponent trembles & Farina and Sandholm (2021)~\cite{Farina21:Equilibrium}\\
Quasi-proper (QPropE) & Properness-style payoff ranking with optimal own continuation & Hansen and Lund (2021)~\cite{Hansen21:Computational}\\
Observable perfect (OPE) & Robustness to publicly observable action-probability trembles & Ganzfried (2023)~\cite{Ganzfried23:Observable}\\
\textbf{Extensive-form proper (EFPropE)} & \textbf{Payoff-ranked local agent actions; other behavior fixed} & \textbf{Theorem~\ref{thm:main} (this work)}\\
\bottomrule
\end{tabular}
\caption{Polynomial-time computability of equilibrium refinements in explicitly represented two-player zero-sum extensive-form games with perfect recall. The last row is the result developed in this paper.}
\label{tab:refinements}
\end{table*}

\section{Introduction}

In sequential games, a Nash equilibrium may prescribe arbitrary behavior
after histories that are assigned probability zero. Trembling-hand
refinements address this weakness by considering fully mixed strategies
and requiring increasingly severe penalties for worse decisions.
Myerson's proper equilibrium~\cite{Myerson78:Refinements} formalizes the principle
that mistakes with strictly lower payoffs should be asymptotically less
likely. Selten~\cite{Selten75:Reexamination} defined extensive-form perfection using
the agent normal form. For extensive-form games, a key modeling choice
is \emph{where} payoff comparisons take place: across complete
contingent plans or separately at information sets.

This paper addresses the latter notion, namely proper equilibrium of
the \emph{agent normal form}. Each information set is represented by an
agent with the corresponding original player's payoff, and comparisons
between the agent's actions hold every other agent's behavior fixed.
The other agents include those representing the same original player
at later information sets. In this sense the definition is a local
refinement of behavioral strategies, rather than properness among
complete normal-form plans. This definition is classical: see van Damme~\cite[Section~6.5]{vanDamme83:Refinements}
and Kohlberg and Mertens~\cite[Section~2.2]{Kohlberg86:Strategic}.
In particular, extensive-form proper equilibrium refines
extensive-form perfect equilibrium. When every information set is
reachable under completely mixed play, the latter also refines
sequential equilibrium~\cite{Kohlberg86:Strategic}. Our
contribution is \emph{computing} the former, not defining it or proving
these inclusions.

A number of other Nash equilibrium refinements can also be computed
in polynomial time for two-player zero-sum game trees. However, the
precise form of the trembling constraints matters: normal-form
properness ranks complete contingent plans, whereas the concept
considered here ranks actions separately at each information set.
Section~\ref{sec:related} compares the applicable complexity results;
Section~\ref{sec:example} gives a zero-sum example separating the
normal-form and agent-form properness requirements.
Other recent work by Hou et al.~\cite{Hou26:Characterization} also employs a
sequence-form approach, but addresses normal-form properness.

This model treats each player's local decisions as separately
perturbed choices with that player's payoff. Anticipating one's
own future mistakes can affect an earlier decision even when the
mistakes vanish. It also carries known conceptual limitations:
extensive-form properness inherits the admissibility concerns discussed
by Hansen and Lund~\cite{Hansen21:Computational} for general games.
Our motivation is to establish exact tractability for this classical
model of payoff-ranked local mistakes; the theorem makes no claim of
universal superiority over the other refinements.

\paragraph{Contributions.}
We establish that an extensive-form proper equilibrium, as defined
above, can be computed \emph{exactly} in time polynomial in the
explicit game-tree input size. The proof combines three ingredients:
(i) a geometric permutahedral perturbation that enforces local
properness at constrained best responses;
(ii) a polynomial-size sequence-form extended formulation using scaled
doubly stochastic matrices; and
(iii) an exact parametric-LP procedure that selects a feasible basis
stable for all sufficiently small positive perturbations and computes
the limit of every behavioral distribution. The third ingredient is
important even when the limit of a realization plan is easy to find,
since its conditional behavior may be undefined on unreached sequences.

\section{Polynomial-Time Algorithms for Related Refinements}
\label{sec:related}

Table~\ref{tab:refinements} summarizes representative exact
polynomial-time results for finite two-player zero-sum extensive-form
games with perfect recall. The complexity measure is the size of the
explicit game-tree representation, rather than the potentially much
larger strategic normal form. The table is a comparison of
\emph{computational results}, not a hierarchy of refinements:
these notions can impose different, sometimes incompatible, criteria.

The distinction between normal-form and extensive-form perfection is
not simply terminological. The former perturbs probabilities of
complete contingent plans, whereas the latter perturbs local actions.
Miltersen and S{\o}rensen~\cite{Miltersen08:Fast} give a
polynomial-time algorithm, based on a sequence of linear programs,
for computing a normal-form proper equilibrium in behavior strategies
of a two-player zero-sum extensive-form game with perfect recall.
Zhang et al.~\cite{Zhang26:Complexity} give an alternative polynomial-time
algorithm for normal-form proper equilibrium in this setting.
Normal-form proper equilibrium ranks \emph{plans}, whereas the classical agent-normal-form
properness criterion~\eqref{eq:proper} ranks local actions while
holding the player's continuation behavior fixed. Van Damme's
result connecting \emph{normal-form} proper equilibrium to
quasi-perfect equilibrium~\cite{vanDamme84:Relation} does not identify it
with agent-normal-form properness. Quasi-perfect equilibrium evaluates
an action without attributing future mistakes to the acting player;
quasi-proper equilibrium adds payoff-based rankings to this
perspective~\cite{Miltersen10:Computing,Hansen21:Computational}.
The example in Section~\ref{sec:example} shows that a normal-form
proper or quasi-proper equilibrium need not be agent-form proper, even
in a zero-sum game. Thus those output guarantees alone do not imply
Theorem~\ref{thm:main}.

Two other refinements specifically motivated by deploying agents
against imperfect opponents are one-sided quasi-perfect equilibrium,
which treats one player as perfectly rational~\cite{Farina21:Equilibrium},
and \emph{observable perfect equilibrium} (OPE), which focuses on
publicly observable action-probability mistakes instead of perturbing
all potentially unobserved actions~\cite{Ganzfried23:Observable}. Ganzfried
also gives a polynomial-time LP-based algorithm for computing OPE in
two-player zero-sum games. The one-sided and observation-conditioned
notions are included because they broaden the set of known tractable
refinements, not because they coincide with agent-normal-form
properness. In particular, the comparison does not claim any
unproved implication between OPE and the properness notion studied here.

The closest perturbation construction is that of Hansen and
Lund~\cite{Hansen21:Computational}, who already combine local
permutahedra, sequence form, and compact extended formulations.
Their small vertex coordinates are depth-dependent powers of $\eps$
in realization space, with the remaining coordinate absorbing the
predecessor weight. Here every vertex coordinate instead scales with
that weight. This homogeneity allows an earlier local distribution
to change while all continuation behaviors stay fixed, which is the
deviation used in Lemma~\ref{lem:saddle}.

Exact stable-basis computation is also an established tool:
Farina, Gatti, and Sandholm~\cite{Farina18:Practical} develop it
for parametric LPs arising in EFPE and QPE computation. We give a
self-contained version for the present feasibility system and retain
the rational functions needed to recover behavior at vanishing
realization weights. The contribution is the local properness
reduction and its exact sequence-form implementation; neither the
permutahedral perturbation nor the general stable-basis principle
is introduced here.

\section{Game Model and Properness}
\label{sec:model}

Let $\Gamma$ be a finite, explicitly represented two-player
extensive-form game with perfect recall. Each terminal history $z$
has payoffs $u_2(z)=-u_1(z)$, and terminal payoffs and chance
probabilities are rational numbers encoded in binary. Chance
probabilities are allowed to be zero. Let $L$ be the total binary input
length. Player $i\in\{1,2\}$ has information sets $\calH_i$;
for $h\in\calH_i$, its available actions are $A(h)$, with
$m_h=|A(h)|\ge1$. A behavioral profile $b=(b_1,b_2)$ consists of
local probability vectors $b_h\in\Delta(A(h))$. It is \emph{completely
mixed} when $b_h(a)>0$ for every $h,a$.

For $h\in\calH_i$ and $a\in A(h)$, let $b^{h\leftarrow a}$
replace $b_h$ by the pure action $a$, leaving every other local
behavioral distribution unchanged. Let $U_i$ be player $i$'s
expected payoff, including chance, and define the unconditional
agent action value
\begin{equation}
  V_h(a;b)=U_i\bigl(b^{h\leftarrow a}\bigr).
  \label{eq:value}
\end{equation}
Notice that the acting player's continuation behavior following
$h$ is held fixed in this comparison.

\begin{definition}[Extensive-form $\eps$-properness]
\label{def:proper}
For $0<\eps<1$, a behavioral profile $b$ is \emph{extensive-form
$\eps$-proper} if it is completely mixed and, for all $i$,
$h\in\calH_i$, and $a,c\in A(h)$,
\begin{equation}
 V_h(a;b)<V_h(c;b)\quad\Longrightarrow\quad
 b_h(a)\le\eps b_h(c).
 \label{eq:proper}
\end{equation}
A profile $b^*$ is an \emph{extensive-form proper equilibrium} if
$b^{(k)}\to b^*$ for some extensive-form $\eps_k$-proper profiles
$b^{(k)}$ with $\eps_k\downarrow0$.
\end{definition}

\begin{theorem}\label{thm:main}
For every finite, explicitly represented two-player zero-sum
extensive-form game with perfect recall and rational payoffs and
chance probabilities, an extensive-form proper equilibrium can be
computed deterministically in time polynomial in $L$.
The algorithm outputs exact rational behavioral probabilities at
\emph{every} information set, with total encoding length polynomial
in $L$. It can also output a rational $\eps_0>0$ and rational
functions $b_h(a;\eps)$ of polynomial total encoding length that
form an extensive-form $\eps$-proper profile for every
$0<\eps\le\eps_0$ and converge to the reported equilibrium.
\end{theorem}

This is Myerson's properness condition applied to the agent normal
form. If $h$ has positive reach probability, the ordering in
\eqref{eq:value} agrees with the corresponding conditional payoff
ordering. If chance makes $h$ impossible to reach, all its
unconditional action values coincide; nevertheless, the algorithm
will specify behavior at $h$. No conditional-rationality claim is
made at such chance-unreachable information sets. The definition is not properness in
the strategic normal form, and it does not allow the player to
re-optimize continuation actions as in quasi-properness.

\section{Geometric Local Perturbations}
\label{sec:geometry}
For $m\ge1$ let
\begin{equation}
 S_m(\eps)=\sum_{k=0}^{m-1}\eps^k,\qquad
 w_m(\eps)=\frac{(1,\eps,\ldots,\eps^{m-1})}{S_m(\eps)},
 \label{eq:weights}
\end{equation}
and let $\mathfrak S_m$ denote the coordinate permutations.
Define the geometric permutahedron
\begin{equation}
 P_m(\eps)=\conv\{\pi w_m(\eps):\pi\in\mathfrak S_m\}.
 \label{eq:permutahedron}
\end{equation}
The geometric properness perturbation is related to that of
Kohlberg and Mertens~\cite{Kohlberg86:Strategic}; see also
S\o rensen~\cite{Sorensen12:Computing}. Every point of $P_m(\eps)$ is
fully mixed: each coordinate is at least
$\eps^{m-1}/S_m(\eps)>0$.

\begin{lemma}[Local ranking]\label{lem:ranking}
Let $v\in\R^m$. Every maximizer $q$ of $v^\top p$ over
$P_m(\eps)$ satisfies $q_a\le\eps q_c$ whenever $v_a<v_c$.
\end{lemma}
\begin{proof}
At a maximizing permutation vertex, $v_a<v_c$ requires
$p_a<p_c$: otherwise swapping the distinct weights at $a,c$
strictly increases the objective. The vertex coordinates have the
form $p_a=\eps^s/S_m$, $p_c=\eps^t/S_m$ for integers $s>t$;
hence $p_a\le\eps p_c$. Every maximizing point is a convex
combination of maximizing vertices, and this inequality is linear.
For $m=1$ there are no action comparisons.
\end{proof}

For example, with three strictly ranked actions $v_1>v_2>v_3$,
the unique local optimizer is
$w_3(\eps)=(1,\eps,\eps^2)/S_3(\eps)$, and each inferior
action is played exactly $\eps$ times as often as the next one.
Ties permit convex combinations of maximizing permutations without
imposing an arbitrary ordering on equally valued actions.

\begin{lemma}[Vanishing perturbation]\label{lem:density}
Every probability vector $q\in\R^m$ is the limit of points
$q(\eps)\in P_m(\eps)$ as $\eps\downarrow0$.
\end{lemma}
\begin{proof}
For each action $a$, choose a permutation vertex with its largest
coordinate at $a$. This vertex tends to the unit vector $e_a$.
Take the convex combination of these vertices with coefficients
$q_a$.
\end{proof}

\subsection{A Separating Example and Its Behavioral Limit}
\label{sec:example}

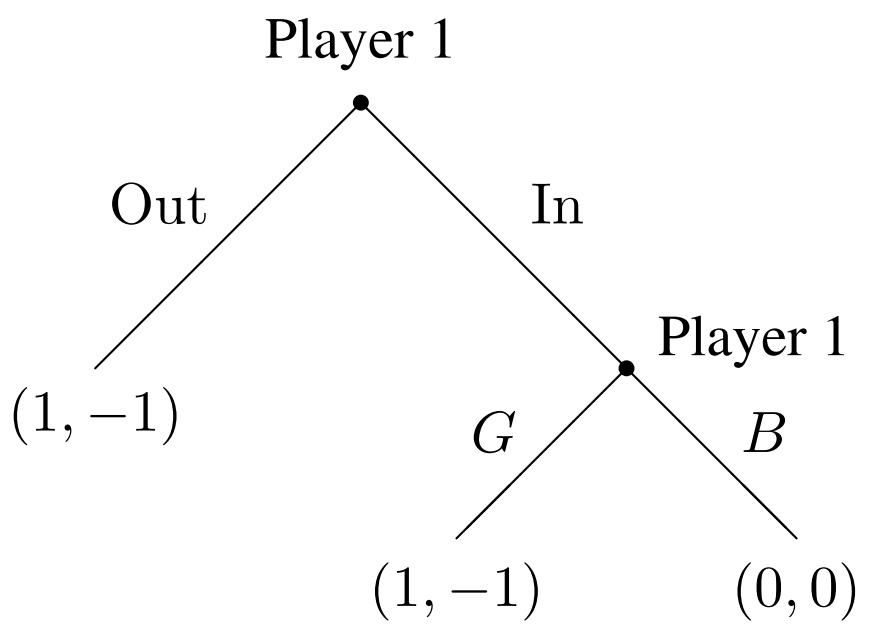
\begin{figure}[t]
\centering
\setlength{\unitlength}{1pt}
\begin{picture}(220,122)
\put(100,104){\circle*{3}}
\put(100,115){\makebox(0,0){Player 1}}
\put(100,104){\line(-1,-1){50}}
\put(100,104){\line(1,-1){50}}
\put(62,85){\makebox(0,0){$\mathrm{Out}$}}
\put(137,85){\makebox(0,0){$\mathrm{In}$}}
\put(50,45){\makebox(0,0){$(1,-1)$}}
\put(150,54){\circle*{3}}
\put(174,59){\makebox(0,0){Player 1}}
\put(150,54){\line(-1,-1){32}}
\put(150,54){\line(1,-1){32}}
\put(125,42){\makebox(0,0){$G$}}
\put(176,42){\makebox(0,0){$B$}}
\put(118,12){\makebox(0,0){$(1,-1)$}}
\put(182,12){\makebox(0,0){$(0,0)$}}
\end{picture}
\caption{A player's own future tremble distinguishes two actions
with equal optimal continuation payoffs. Payoffs are $(u_1,u_2)$.}
\label{fig:continuation}
\end{figure}

Consider Figure~\ref{fig:continuation}, a zero-sum version of the
one-player tree in Kohlberg and Mertens~\cite[Appendix~A,
Figure~15]{Kohlberg86:Strategic}. Player~1 chooses $\mathrm{Out}$,
receiving~1, or $\mathrm{In}$, after which the same player chooses
$G$ or $B$, receiving~1 or~0. Player~2 has no decisions and
receives the negative payoff. Write $p=b(\mathrm{In})$ and
$q=b(B\mid\mathrm{In})$.

At any completely mixed profile, $q>0$, so choosing
$\mathrm{In}$ yields $1-q<1$, the payoff from $\mathrm{Out}$.
Agent-form $\eps$-properness requires $p\le\eps(1-p)$.
At the second decision, $p>0$ makes $G$ strictly better than $B$,
so also $q\le\eps(1-q)$. Every agent-form proper limit therefore
chooses $\mathrm{Out}$ initially and $G$ after $\mathrm{In}$.

In the strategic normal form the plans
$(\mathrm{Out},G)$, $(\mathrm{Out},B)$, $(\mathrm{In},G)$,
and $(\mathrm{In},B)$ have payoffs $1,1,1,0$.
For $0<\eta<1$, assign them probabilities proportional to
$(\eta,\eta,1,\eta^2)$. This is $\eta$-proper and converges
to $(\mathrm{In},G)$, which is not agent-form proper.
The same limiting behavior is quasi-proper: with optimal own
continuation, $\mathrm{Out}$ and $\mathrm{In}$ both yield~1.
Indeed, $p=1/(1+\eta)$ and $q=\eta/(1+\eta)$ form a
completely mixed $\eta$-quasi-proper family.

For our perturbation, $P_2(\eps)$ puts each action probability
in $[\eps/(1+\eps),1/(1+\eps)]$. Maximizing the payoff
$1-pq$ gives $p=q=\eps/(1+\eps)$, with realization weights
\[
 r(\mathrm{In})=\frac{\eps}{1+\eps},\qquad
 r(\mathrm{In},G)=\frac{\eps}{(1+\eps)^2}.
\]
Both weights vanish, but their ratio converges to~1. Separate
limits cannot recover the required continuation $G$, motivating
the limits of behavioral ratios in Section~\ref{sec:limit}.

\section{A Polynomial-Size Sequence-Form Polytope}
\label{sec:extended}
For player $i$, let $\calQ_i$ be the set of sequences of the
player's information-set/action pairs, including the empty sequence
$\varnothing$. By perfect recall, every $h\in\calH_i$ has a
well-defined predecessor sequence $\sigma(h)$; its extension by
$a\in A(h)$ is denoted $\sigma(h)a$.
The realization plan $r_i$ satisfies
\begin{equation}
 r_i(\varnothing)=1,\quad
 \sum_{a\in A(h)}r_i(\sigma(h)a)=r_i(\sigma(h)),\quad r_i\ge0.
 \label{eq:flow}
\end{equation}
For a positive predecessor realization weight, the behavioral
probability is $b_h(a)=r_i(\sigma(h)a)/r_i(\sigma(h))$.
We require $b_h\in P_{m_h}(\eps)$, without enumerating its $m_h!$
vertices.

Fix $h$, and abbreviate $m=m_h$, $p=r_i(\sigma(h))$,
$t_a=r_i(\sigma(h)a)$. Introduce $m^2$ nonnegative variables
$Z^h_{ak}$, with $a\in A(h)$ and $0\le k<m$, and require
\begin{subequations}\label{eq:localEF}
\begin{align}
 \sum_{k=0}^{m-1} Z^h_{ak}&=p
       && (a\in A(h)), \label{eq:localEFa}\\
 \sum_{a\in A(h)} Z^h_{ak}&=p
       && (0\le k<m), \label{eq:localEFb}\\
 S_m(\eps)t_a&=\sum_{k=0}^{m-1}\eps^k Z^h_{ak}
       && (a\in A(h)), \label{eq:localEFc}\\
 Z^h_{ak}&\ge0. \label{eq:localEFd}
\end{align}
\end{subequations}

\begin{lemma}[Exact extended formulation]\label{lem:formulation}
For $p>0$, constraints~\eqref{eq:localEF} hold for some $Z^h$
if and only if $t/p\in P_m(\eps)$. When $p=0$, they force
$t=0$ and $Z^h=0$. In all cases they imply
\begin{equation}
 \sum_a t_a=p,\qquad
 t_a\ge \frac{\eps^{m-1}}{S_m(\eps)}p.
 \label{eq:lowerbound}
\end{equation}
\end{lemma}
\begin{proof}
For $p>0$, the matrix $D=Z^h/p$ is doubly stochastic.
By the Birkhoff--von Neumann theorem, it is a convex combination of
permutation matrices, and~\eqref{eq:localEFc} states
$t/p=Dw_m(\eps)$. The converse follows by scaling a doubly
stochastic representation by $p$. If $p=0$, nonnegativity and
the row sums force $Z^h=0$, hence $t=0$. Summing
\eqref{eq:localEFc} over $a$ gives the first identity in
\eqref{eq:lowerbound}; its second inequality follows from the
lower bound $\eps^k\ge\eps^{m-1}$ and~\eqref{eq:localEFa}.
\end{proof}

Let $\mathcal R_i(\eps)$ be the extended set of $(r_i,Z)$
with $r_i(\varnothing)=1$, nonnegativity, and
\eqref{eq:localEF} imposed at every information set.
The flow constraints follow from Lemma~\ref{lem:formulation}.
Its projection onto the realization coordinates is \emph{exactly}
the set of plans induced by behaviors in the local permutahedra.
Indeed, the lemma supplies a local matrix in both directions, and
\eqref{eq:lowerbound} inductively implies that every realization
coordinate is strictly positive. The uniform local distributions
establish nonemptiness. All realization and auxiliary coordinates
lie in $[0,1]$, so the polytope is compact.
There are $1+\sum_h m_h$ realization coordinates and
$\sum_h m_h^2$ auxiliary variables. All equality coefficients are
polynomials in $\eps$ with small integer coefficients. The block
for $h$ has degree at most $m_h-1$, and the normalization has
degree zero. The encoding size is polynomial in $L$.

\section{From Saddle Points to Properness}
\label{sec:saddle}
Write the two extended polytopes in standard form,
\begin{equation}
\begin{aligned}
 \mathcal X_\eps&=\{x\ge0:E(\eps)x=e\},\\
 \mathcal Y_\eps&=\{y\ge0:F(\eps)y=f\}.
\end{aligned}
 \label{eq:polytopes}
\end{equation}
The vectors include realization and auxiliary variables. From
sequence form~\cite{Koller96:Efficient}, a rational payoff
matrix $A$, padded by zero rows and columns for auxiliary variables,
satisfies $U_1=x^\top Ay$. Its nonzero block has entries
\begin{equation}
 A_{\sigma,\tau}=\sum_{\substack{z\text{ terminal}:\\
   \sigma_1(z)=\sigma,\ \sigma_2(z)=\tau}}
 \pi_c(z)u_1(z),
 \label{eq:payoff}
\end{equation}
where $\pi_c(z)$ is the product of chance probabilities on the
path to $z$. Explicit tree representation and rational input
ensure that $A$ is polynomial-time constructible with polynomial
bit length. Consider the zero-sum saddle-point problem
\begin{equation}
 \max_{x\in\mathcal X_\eps}\min_{y\in\mathcal Y_\eps}
 x^\top Ay.
 \label{eq:saddle}
\end{equation}
For each $0<\eps<1$, the strategy sets are nonempty, compact,
and convex, and the payoff is bilinear. The minimax theorem
therefore guarantees a saddle point, characterized by
\begin{equation}
 (x')^\top Ay\le x^\top Ay\le x^\top Ay'
 \quad(\forall x',y'\text{ feasible}).
 \label{eq:saddleineq}
\end{equation}

\begin{lemma}[Saddle-point properness]\label{lem:saddle}
Every saddle point of~\eqref{eq:saddle} induces an
extensive-form $\eps$-proper behavioral profile.
\end{lemma}
\begin{proof}
Lemma~\ref{lem:formulation} ensures that the induced profile $b$ is
completely mixed. Fix $h\in\calH_i$ and any $q\in P_{m_h}(\eps)$.
Let $b'=b^{h\leftarrow q}$, leaving every other local distribution
fixed, and let $r'_i$ be its induced realization plan. Each $b'_g$
still lies in its local permutahedron. For every $g\in\calH_i$,
choose a doubly stochastic matrix $D^g$ with
$b'_g=D^g w_{m_g}(\eps)$ and set
$Z'^g=r'_i(\sigma(g))D^g$.
Lemma~\ref{lem:formulation} shows that $(r'_i,Z')$ is feasible.
Thus realization weights can change without changing any
continuation behavior.
By perfect recall, a path never visits the same information set
twice, so payoff is affine in the distribution at $h$:
\begin{equation}
 U_i\bigl(b^{h\leftarrow q}\bigr)
   =\sum_{a\in A(h)}q_a V_h(a;b).
 \label{eq:affine}
\end{equation}
For player~1 the left saddle inequality, and for player~2
the right saddle inequality together with $U_2=-U_1$,
imply that $b_h$ maximizes~\eqref{eq:affine} over the local
permutahedron. Lemma~\ref{lem:ranking} gives exactly
\eqref{eq:proper} at $h$.
\end{proof}

A saddle point is characterized by linear primal--dual
feasibility. Introduce unrestricted dual vectors $\alpha,\beta$:
\begin{subequations}\label{eq:primaldual}
\begin{align}
 E(\eps)x&=e, &x&\ge0,\\
 F(\eps)y&=f, &y&\ge0,\\
 E(\eps)^\top\alpha&\ge Ay,\\
 F(\eps)^\top\beta&\le A^\top x,\\
 e^\top\alpha&=f^\top\beta.
\end{align}
\end{subequations}
For any feasible solution to~\eqref{eq:primaldual},
nonnegativity and the dual inequalities give
\[
 x^\top Ay\le e^\top\alpha=f^\top\beta\le x^\top Ay,
\]
so both dual bounds equal the current payoff. For any feasible
$x',y'$, weak duality gives $(x')^\top Ay\le e^\top\alpha$
and $f^\top\beta\le x^\top Ay'$, proving~\eqref{eq:saddleineq}.
Conversely, at any saddle point, LP strong duality provides
$\alpha,\beta$ with the common value $x^\top Ay$.
Thus the projection of~\eqref{eq:primaldual} onto $(x,y)$ is
\emph{precisely} the saddle-point set.
Splitting unrestricted variables into nonnegative parts, adding
slacks, and clearing rational denominators yields
\begin{equation}
 M(\eps)z=d(\eps),\qquad z\ge0,
 \label{eq:standard}
\end{equation}
with $M,d$ over $\Z[\eps]$. Dimensions, entry degrees, and
coefficient bit lengths are polynomial in $L$. The system is
feasible for all $0<\eps<1$.

\section{Exact Polynomial-Time Limit Extraction}
\label{sec:limit}
The final step is not simply to solve~\eqref{eq:standard} at a
small numerical value of $\eps$: the desired equilibrium is a
\emph{limit}, and an arbitrary perturbation does not certify it.
Stable-basis methods for trembling LPs have been developed previously
for EFPE and QPE~\cite{Farina18:Practical}. Here we use a
self-contained bit-complexity argument to find a basic feasible
solution that remains valid throughout $(0,\eps_0]$, then take
the \emph{behavioral ratios} before passing to the limit.

\subsection{Uniform Sign Stability}
Let $R\ge1$ be the number of equality rows in
\eqref{eq:standard}, $D$ the maximum degree of any entry of
$[M(\eps)\mid d(\eps)]$, and $C\ge1$ an integer upper bound
on the absolute value of every entry coefficient. Define
\begin{equation}
 H=R!\bigl((D+1)C\bigr)^R,\qquad
 \eps_0=\frac{1}{2H+2}.
 \label{eq:eps0}
\end{equation}

\begin{lemma}[Signs of all minors]\label{lem:sign}
Every square minor of $[M(\eps)\mid d(\eps)]$ is either the
zero polynomial or has constant, nonzero sign throughout
$0<\eps\le\eps_0$. The rational $\eps_0$ is computable in
polynomial time and has polynomial binary encoding length.
\end{lemma}
\begin{proof}
An order-$r$ determinant ($r\le R$) sums at most $r!$
products of $r$ polynomial entries. Each coefficient therefore
has absolute value at most
$r!((D+1)C)^r\le H$.
For a nonzero integer polynomial
$p(\eps)=\eps^k(a_k+a_{k+1}\eps+\cdots)$,
where $a_k\ne0$, we have $|a_k|\ge1$ and
\begin{equation}
 \left|\sum_{j>k}a_j\eps^{j-k}\right|
 \le\frac{H\eps}{1-\eps}<\frac12
 \quad(0<\eps\le\eps_0).
 \label{eq:tail}
\end{equation}
Its sign is therefore the sign of $a_k$, including at
$\eps_0$. Furthermore
$\log_2 H=O(R\log(R+1)+R\log(D+1)+R\log C)$,
which is polynomial in $L$.
\end{proof}

\subsection{One Feasible Basis for All Small $\eps$}
Evaluate~\eqref{eq:standard} at the exact rational $\eps_0$,
and let $n$ be its number of variables. Substitution increases
coefficient bit lengths by at most a polynomial amount: a degree-$D$
entry has denominator dividing $(2H+2)^D$. Thus an exact rational
LP algorithm applies~\cite{Schrijver86:Theory}.

To obtain a vertex, first minimize $\mathbf1^\top z$ over
the nonempty standard-form polyhedron. This objective is bounded
below, and its optimal face is compact because $z\ge0$ and
$\mathbf1^\top z$ is fixed. Successively minimize each coordinate
over the preceding optimal face. After at most $n+1$ LP calls,
all coordinates are fixed. The resulting singleton is a face,
hence a vertex, of the original polyhedron.

Every intermediate optimal face is also a face of the original
polyhedron and contains one of its vertices. Each newly fixed
value is therefore a coordinate of an original vertex, or its
coordinate sum for the first objective. Rational-vertex determinant
bounds give polynomial encoding length for all these values in
terms of the original LP data; the bounds do not compound across
the successive calls.

At a vertex the columns indexed by positive coordinates are
linearly independent: a dependence would permit a small feasible
move in both directions. After removing redundant rows, extend
these columns by Gaussian elimination to a basis. Its additional
basic coordinates are zero. This handles degeneracy without
assuming a polynomial pivot bound for the simplex method.

Let $r=\rank M(\eps_0)$, and choose $r$ independent equality
rows $I$ and $r$ basic columns $J$. Put
\begin{equation}
 B(\eps)=M_{I,J}(\eps),\quad
 z_J(\eps)=B(\eps)^{-1}d_I(\eps),\quad
 z_{J^c}(\eps)=0.
 \label{eq:basis}
\end{equation}
The normalization of the empty sequences ensures $r\ge1$.

\begin{lemma}[Persistent feasible basis]\label{lem:stable}
The rational-function vector $z(\eps)$ defined by
\eqref{eq:basis} satisfies~\eqref{eq:standard} for every
$0<\eps\le\eps_0$.
\end{lemma}
\begin{proof}
Because the determinant $\Delta(\eps)=\det B(\eps)$ is
nonzero at $\eps_0$, Lemma~\ref{lem:sign} keeps it nonzero
throughout the interval, so the rows indexed by $I$ stay
independent. The ranks of $M(\eps)$ and $[M(\eps)\mid d(\eps)]$
remain constant by the same lemma; both equal $r$ at $\eps_0$
by feasibility. Hence the selected $r$ augmented rows span the
augmented row space throughout the interval. Every omitted row,
including its right-hand side, is a linear combination of these
rows. Satisfying the equations indexed by $I$ therefore satisfies
all equations in~\eqref{eq:standard}.
Cramer's rule yields, for each basic coordinate,
\begin{equation}
 z_j(\eps)=\frac{N_j(\eps)}{\Delta(\eps)},
 \label{eq:cramer}
\end{equation}
where $N_j$ is a signed minor of the augmented matrix.
Both $N_j$ and $\Delta$ have stable signs. As each ratio is
nonnegative at $\eps_0$, it is nonnegative throughout the
interval. If a basic coordinate is zero at $\eps_0$, its numerator
must be the zero polynomial, explicitly covering degeneracy.
\end{proof}

The polynomials $\Delta$ and $N_j$ have degree at most $rD$
and coefficients bounded in absolute value by $H$, so their dense
representations have polynomial bit length. Compute them by
evaluating determinants at $rD+1$ distinct integers and
interpolating; one evaluation suffices if $D=0$. Both computations
have polynomial bit complexity, and there are at most $n$ basic
coordinates. Dual variables may diverge as $\eps\downarrow0$,
but only bounded realization coordinates and their ratios
determine the output.

\subsection{Recovering All Behavioral Limits}
For each $\eps\in(0,\eps_0]$, the realization coordinates of
$z(\eps)$ define a saddle point and hence an
$\eps$-proper profile by Lemma~\ref{lem:saddle}.
They are strictly positive, so the corresponding coordinates
are basic. Since the Cramer denominator cancels, for every
$h\in\calH_i$ and $a\in A(h)$,
\begin{equation}
 b_h(a;\eps)=\frac{r_i(\sigma(h)a;\eps)}
                      {r_i(\sigma(h);\eps)}
                =\frac{N_{\sigma(h)a}(\eps)}
                       {N_{\sigma(h)}(\eps)}.
 \label{eq:behavior}
\end{equation}
Both numerator polynomials in the last ratio are nonzero.
For a nonzero polynomial $P$, let $\ord(P)$ be the smallest
exponent with a nonzero coefficient and $\lc(P)$ that
coefficient. Let $k=\ord(N_{\sigma(h)a})$ and
$\ell=\ord(N_{\sigma(h)})$. Because
$0<b_h(a;\eps)\le1$, necessarily $k\ge\ell$, and
\begin{equation}
 b_h^*(a)=\lim_{\eps\downarrow0}b_h(a;\eps)
 =\begin{cases}
 0, & k>\ell,\\[2pt]
 \displaystyle\frac{\lc(N_{\sigma(h)a})}
                    {\lc(N_{\sigma(h)})}, & k=\ell.
 \end{cases}
 \label{eq:limit}
\end{equation}
Scanning coefficient lists computes these exact rational limits
in polynomial time. Passing to the limit in each local
normalization shows that they sum to one at every information set.

\paragraph{Why Conditional Limits Are Essential.}
If $r_i(\sigma(h);\eps)\to0$, the limit of the realization
plan alone does not determine $b_h$. Arbitrarily filling in
behavior at $h$ would not in general preserve properness.
Equation~\eqref{eq:limit} instead keeps the relative leading
orders of two vanishing realization weights.

\section{Algorithm and Consequences}
\label{sec:algorithm}

\begin{algorithm}[t]
\caption{Exact extensive-form proper equilibrium}
\label{alg:main}
\noindent\textbf{Input}: Explicit two-player zero-sum perfect-recall
game with rational data.\par
\noindent\textbf{Output}: Exact proper behavioral profile $b^*$.\par
\begin{algorithmic}[1]
\STATE Construct $A$ and the local scaled doubly stochastic
       formulations~\eqref{eq:localEF}.
\STATE Build the saddle-point system~\eqref{eq:primaldual}; convert
       it to~\eqref{eq:standard}.
\STATE Compute $R,D,C,H$ and $\eps_0$ by~\eqref{eq:eps0}.
\STATE Find an exact basic feasible point at $\eps_0$ and a
       corresponding row/column basis $(I,J)$.
\STATE Obtain the polynomial Cramer numerators for the
       realization coordinates.
\STATE \textbf{return} all local ratios' limits using~\eqref{eq:limit}.
\end{algorithmic}
\end{algorithm}

\begin{proof}[Proof of Theorem~\ref{thm:main}]
Lemmas~\ref{lem:formulation} and~\ref{lem:saddle} imply that
each fixed-parameter feasible saddle point produces an
$\eps$-proper profile. Lemmas~\ref{lem:sign} and~\ref{lem:stable}
produce a feasible rational-function saddle point on the entire
interval $(0,\eps_0]$, and~\eqref{eq:limit} computes its exact
behavioral limit $b^*$. Taking, for example,
$\eps_k=\eps_0/(k+1)$ proves properness by
Definition~\ref{def:proper}. All dimensions, determinant degrees,
and rational arithmetic bit lengths are polynomial in $L$,
establishing the claimed running time and output size.

For completeness, $b^*$ is a Nash equilibrium \emph{of the
original two-player game}, not merely an equilibrium against
individual agents. Fix any behavioral deviation $\widehat b_i$.
By Lemma~\ref{lem:density}, approximate each of its local
distributions by a point in the corresponding
$P_{m_h}(\eps)$, obtaining feasible
$\widehat b_i(\eps)\to\widehat b_i$. The saddle-point property
implies, for every small $\eps$,
\begin{equation}
 U_i\bigl(b_i(\eps),b_{-i}(\eps)\bigr)
 \ge U_i\bigl(\widehat b_i(\eps),b_{-i}(\eps)\bigr).
 \label{eq:nashlimit}
\end{equation}
Continuity of expected payoffs gives
$U_i(b^*)\ge U_i(\widehat b_i,b_{-i}^*)$.
Perfect recall gives the usual equivalence for mixed-strategy
deviations, completing the proof.
\end{proof}

\paragraph{Size and Arithmetic Bounds.}
Put $s=\sum_i|\calQ_i|$ and
$t=\sum_i\sum_{h\in\calH_i}m_h^2$. The standard-form system
has $n,R=O(s+t)$, including dual variables and slacks; both
$s,t$ are polynomial in $L$. Clearing denominators preserves
polynomial coefficient bit length. Entry degrees are bounded by
the largest local action count, or zero if there are no information
sets. At most $n+1$ exact LP calls select the basis. At most $n+1$
determinant polynomials, of degree at most $RD$, require $RD+1$
evaluations each. Evaluation points, determinant values, and
interpolated coefficients have polynomial bit length, as do the
rational-function witness and exact behavioral limits in
Theorem~\ref{thm:main}.

\paragraph{Scope and Limitations.}
The result establishes polynomial \emph{bit} complexity for an
explicit game tree, not strongly polynomial time or practical
efficiency at $\eps_0$. It assumes zero-sum payoffs and perfect
recall, but no genericity or nondegeneracy.
The output is one exact agent-form proper equilibrium, hence also
agent-form perfect. If every information set is reachable under
completely mixed play, it is also sequential~\cite{Kohlberg86:Strategic}.
At information sets made unreachable by chance, unconditional agent
properness imposes no conditional optimality requirement.
The method neither enumerates all proper equilibria nor establishes
complexity results for general-sum games.
Although first-order methods can scale better for other
refinements~\cite{Kroer17:Smoothing}, efficient implementation
of this exact construction remains future work.

\section*{Acknowledgments}
ChatGPT (OpenAI) played a substantial role in developing the main theorem and its proof. ChatGPT and Codex (OpenAI) also assisted with checking and clarifying the mathematical arguments and preparing the manuscript. The author takes responsibility for the final content.

\apptocmd{\thebibliography}{\setlength{\itemsep}{2pt}}{}{}
\bibliographystyle{plain}
\bibliography{C://FromBackup/Research/refs/dairefs}
\end{document}